\documentstyle [12pt]{article}
\begin{document}
 
\thispagestyle{empty}
 
\setlength{\unitlength}{1in}
\begin{center}
{\Large{\bf Bose-Einstein Condensation of Atoms in a Trap}}
\end{center}
\baselineskip 0.2in
\begin{center}
\  \\
\  \\
{\bf T.T. Chou}$^1$, {\bf Chen Ning Yang}$^2$ {\bf  and L.H. Yu}$^3$\\
\ \\
$^1${\small\em{Department of Physics, University of Georgia, Athens, 
Georgia 30602}} \\
$^2${\small\em{Institute for Theoretical Physics, State University of New
York}}, \\
{\small\em {Stony Brook, New York 11794}} \\
 
$^3${\small\em{National Synchrotron Light Source, Brookhaven National
Laboratory, Upton}}, \\
{\small\em {New York 11973}}\\
\ \\
\ \\
\end{center}
\begin{abstract}
\normalsize
We point out that the local density approximation (LDA) of Oliva is
an adaptation of the Thomas-Fermi method, and is a good approximation 
when $\varepsilon = \hbar\omega/kT <<1$.  For the case of scattering 
length $a > 0$, the LDA leads to a quantitative result (14') easily 
checked by experiments.  Critical remarks are made about the physics 
of the many body problem in terms of the scattering length $a$.
\end{abstract}

\newpage
\baselineskip 0.30in
 
Bose-Einstein condensation (BEC) of free particles was a great contribution
to physics which provides one more example of the awesome and daring 
insight
of Einstein.  For some sixty years, experimental observation of BEC was
considered hopeless.  Now through a series of ingenious developments BEC 
has
finally been observed [1-3] for trapped bosons.  We would like to discuss
here BEC of such trapped particles with and without mutual interactions [4]
.
First some preliminaries.
 
Four lengths are involved in harmonic traps.  They are:
\begin{equation}
{L_1} = ({2\pi}m\omega^{2}\beta)^{-1/2}, \ \ L_2 =
({\hbar/m}\omega)^{1/2}, \ \ \lambda = (2\pi\hbar^{2}\beta/m)
^{1/2}, \ \ a,  
\end{equation}
where  ${\beta} = 1/kT$, and $L{_1}$ gives the order of the 
classical oscillation amplitude of a particle in the oscillator with 
energy $kT$,  $L{_2}$ is the size of the ground state in the oscillator, 
$\lambda$ the thermal wave length and $a$ the s-wave scattering length.  
 For the recent experiments $\lambda>>|a|$.  We notice also that
\begin{equation}
        {L_1}:{L_2}:{\lambda}=1:{\sqrt{2\pi\varepsilon}}:{2\pi
\varepsilon}, \ \ \varepsilon = \beta\hbar\omega.                           
\end{equation}
 
\noindent {\bf 1. Thomas-Fermi method and the local density approximation 
(LDA), gaseous phase}
 
 For small values of $\varepsilon$ and large number of particles, 
the Thomas-Fermi method for atoms can be adapted to the present 
problem:  We divide space up into cells of volume larger than 
$(L_{2})^{3}$ but smaller than $(L_{1})^{3}$ ,
and consider the potential energy $V$ of the trap to be a constant 
in each cell.  Each cell contains a collection of particles 
with mutual interaction described by a scattering length $a$.  
All particles in a cell have the same external potential $V(r)$ 
per particle.   If $V=0$ this problem for a cell has been studied 
[5,6] in the 1950s.  For the present problem, in each cell we
need only replace the fugacity $z$ of the 1950 result by 
$z \ {\rm exp}(-\beta V)$.
 
This adaptation of the Thomas-Fermi method to the present problem has 
been used in a paper by Oliva (see Ref.[4]).  We shall follow 
his terminology and call it the local density approximation 
(LDA).  In this approximation, in the gaseous phase (i.e., without BEC) 
in each cell, $\rho(r)$ is given [5] by 
\begin{equation}
       \rho(r) = {\lambda^{-3}}{g_{3/2}}(\zeta) 
[1-(4a/{\lambda})g_{1/2}(\zeta)] 
\end{equation}
where 
\begin{equation}
        \zeta = z \ {\rm exp}(-\beta V), \ \ \ g{_\nu}(x) = 
{\sum_{1}^{\infty}}{\ell}^{-\nu}x^{\ell} 
\end{equation}
\noindent and $z$ is the fugacity. \\
To order $a/\lambda$ we can also write (3) as
\begin{equation}
        \rho(r) = \lambda^{-3}g_{3/2}(\xi) 
\end{equation}
\begin{equation}
        \xi = z \ {\rm exp}[-{\beta}V(r) - 4a\lambda^{2}\rho(r)]. 
\end{equation}
 
\noindent {\bf 2. Validity of LDA for case a=0}
 
If $a=0$, LDA [i.e., (5) and (6)] gives immediately an 
explicit expression for $\rho(r)$ provided $z\leq1$.  If we pack 
in more particles, i.e., if we try to increase $z$ beyond 1, since 
$\xi$ cannot further increase at $r=0$, BEC sets in at $r=0$, in the 
same way as the original Einstein description
of BEC in momentum space at $p = 0$. But this case of $a=0$ and
$V = \frac{1}{2} m{\omega^{2}}{r^{2}}$ can be rigorously solved, 
allowing for an understanding of how LDA approaches the rigorous 
result.  This is what we shall do in the present section. 
 
The density of atoms is $\rho(r)$ = $\langle{r|D|r}\rangle$ where D
is the density matrix equal to $z \ e^{-\beta H}(1-z \ e^{-\beta H})^{-1}$
= $\sum_{1}^{\infty} z^{\ell} e^{-\beta H\ell}$. The matrix elements of     
$e^{-\beta H {\ell}}$ is explicitly known from [7].  (Using it we avoid
the tedious process of summing over squares of Hermite polynomials.)
We thus obtain
\begin{equation}
        \rho(r) = \varepsilon^{3/2} \lambda^{-3} \sum_{1}^{\infty}
({\sinh} \ \ell\varepsilon)^{-3/2} \ z^{\ell} \ {\rm exp}
[-\sigma^{2} \ {\tanh}(\ell\varepsilon /2)]
\end{equation}
where $\sigma = r/L_{2}$.  The summand in (7) behaves at large $\ell$
like a geometrical series with the ratio of successive terms equal to 
\begin{equation}
        z_{1} = z \ e^{-3{\varepsilon}/2}
\end{equation}
Thus the summation converges at all $r$ for $z_{1} < 1$ and becomes
divergent at $z_{1} = 1$ for all $r$.  To study how it diverges we
write the summation as ${\sum_{1}^{\infty}}a_{\ell}$ where 
\begin{eqnarray*}
{\ell}n \ a_{\ell} & = & {\ell}({\ell}n \ z) - (3/2) {\ell}n({\sinh} \ 
\ell\varepsilon) - \sigma^2 \ {\tanh}(\ell\varepsilon/2)  \\
& = & {\ell}[{\ell}n \ z - 3 \varepsilon/2] + 
(\frac{3}{2} \ {\ell}n2 - \sigma^2) + C_{\ell}
\end{eqnarray*}
where \ $C_{\ell} \rightarrow 0$ as $\ell \rightarrow \infty$. 
Thus \ $a_{\ell} = 2^{3/2} \ e^{-\sigma^{2}} z^{\ell}_{1} + 2^{3/2} \
e^{-\sigma^{2}} z^{\ell}_{1}(e^{C_{\ell}} - 1)$.\\
Therefore the summation is equal to
\begin{equation}
\sum a_{\ell} = 2^{3/2} \ e^{-\sigma^{2}} \frac{z_1}{1-z_{1}} + 
\ 2^{3/2} \ e^{-\sigma^{2}} \sum z^{\ell}_{1} (e^{C_{\ell}} - 1).
\end{equation}
The first term on the right is the divergent part as $z_{1} \rightarrow$ 1.
We notice happily that it is exactly proportional to $|\psi_{0}(r)|^2$ where 
$\psi_{0}$ is the normalized ground state wave function of the
harmonic oscillator.  Thus (7) becomes
\begin{equation}
\rho(r) = \frac{z_{1}}{1-z_{1}} |\psi_{0}(r)|^{2} + \rho_{n}(r) 
\end{equation}
showing that the BEC is in the ground state.  $\rho_{n}$ is the normal
fluid part of the density function $\rho(r)$.  It is the last term of (9)
multiplied by $\lambda^{-3} \varepsilon^{3/2}$:
\begin{equation}
\rho_{n}(r) = \lambda^{-3}(2\varepsilon)^{3/2} \sum^{\infty}_{1} 
z^{\ell}_{1} \left\{\frac{1}{(1-e^{-2\ell\varepsilon})^{3/2}} \ {\rm exp}
[- \sigma^{2} \tanh(\varepsilon \ell / 2)] - {\rm exp} 
[- \sigma^{2}] \right\}.   
\end{equation}
So far (10) and (11) are exact.  Notice that for large $\ell$, the sum 
in (11) is convergent for $z_{1}=1$, unlike the first term of (10).  Now 
we go to the case of $\varepsilon <<1$.  The contribution to the sum in 
(11) for $\varepsilon\ell>1$ is negligible.  For $\varepsilon\ell<1$, can 
drop the last term in the curly bracket and replace 
$1 - e^{-2\ell\varepsilon}$
by $2\ell\varepsilon$ and $\tanh(\varepsilon\ell/2)$ by 
$\varepsilon\ell/2$, obtaining
\begin{equation}
\rho(r) \cong\frac{z_{1}}{1-z_{1}} |\psi_{0}(r)|^{2} + \lambda^{-3}g_{3/2}
(z_{1}e^{-\frac{1}{2} \beta m \omega^{2}r^{2}}).\\
\end{equation}
The second term here is the result of LDA with the replacement of $z$  
by $z_{1}$.  See (8).  For $\varepsilon<<1$, this replacement creates 
negligible errors.
 
In the gaseous phase, i.e., $1-z_{1} {\approx} O(1)$, the second term 
in (12) dominates and the LDA is good.
 
        In the BEC phase, the first term is the condensate, with            
$1-z_{1} {\approx} O(N^{-1})$.  Thus we can put $z_{1}=1$  in the 
second term, obtaining exactly the result of LDA, except for the 
fact that in the rigorous result the condensate has the spatial
dependence of ${|\psi_{0}(r)|}^{2} = {\rm (const.) \ exp}[-\sigma^{2}]$
while in LDA the condensate has a delta-function dependence on $\vec{r}$. 
 But this is hardly surprising since in LDA each cell has
a linear dimension large compared with $L_{2}$, so that any 
structure of the order of $L_{2}$ is shrunk to a point.  
This fact also means that in LDA the cells should not be chosen to 
be $\leq$ the order $L_{2}$.
 
        To further check the error in (12) we evaluated numerically its 
error divided by the second term on the right for certain cases 
and found that the ratio is generally of the order of 
$\sqrt{\varepsilon}$.  For example, if $\varepsilon = 0.05$, the ratio 
is $< 0.28$, and for $\varepsilon = 0.01$, the ratio is $< 0.12$.
 
        We summarize:  As $\varepsilon \rightarrow 0$, the LDA expression 
 for $\rho(r)$ approaches the rigorous result at every fixed $\sigma_{1}
= r/L_{1}$.
 
        We mention here that with $a = 0$, there is the well-known 
basic symmetry in the Hamiltonian between the coordinate and 
the momentum.  To be precise, if we keep $\hbar$, $\omega$ 
and $\beta$ unchanged, but replace $m$ with                   
${(m\omega^{2})}^{-1}$ and switch $x$ and $p$, the problem is 
unchanged  (for $a = 0$).  Thus we easily obtain the exact momentum
space density distribution  $n(p)$.  The condensate is of course in 
the state ${|\psi_{0}(p)|}^{2}$.
 
\noindent {\bf  3.  Condensate for the case a $>$ 0 in LDA} 
 
 For $a > 0$, the thermal equilibrium in each cell can be studied as 
in [6].  For large enough density, condensation takes place in 
the cell.  The cell then consists of a saturated gaseous 
part with density $\rho_{0} = \lambda^{-3}g_{3/2}(1)$ plus a super
part with density $\rho - \rho_{0} = \rho_{s}$.   
The free energy density then becomes, according to
Eq.(33) of Ref.[6]:
\begin{equation}
f(r) = -kT\lambda^{-3}g_{3/2}(1) + 2a\lambda^{2}kT\rho^{2} - 
a\lambda^{2}kT\rho^{2}_{s}.
\end{equation}                                                              
                  
  From this free energy we obtain the chemical potential which should be
equated with $kT{\ell}n [z \ {\rm exp}(-\beta V)]$  giving
\begin{equation}
kT \ {\ell}n \ z - V = 4\pi a [\rho + 
\rho_{0}]\hbar^{2}/m \ \ \ \ \ \ (r < r_{0}).
\end{equation}
At $r = r_{0}-$, $\rho_{s} = 0$ and $\rho = \rho_{0}$.  Thus \\
\indent \hspace{2cm}  \ \ \ $V(r) + 4\pi a\rho(r)\hbar^{2}/m 
= V(r_{0})+4\pi a \rho_{0} \hbar^{2}/m$. \hfill (14') \\
This simple equation is valid in LDA for any trap potential              
$V(\vec{r})$ and should be {\em testable experimentally}.
 
        Outside of $r_{0}$, (5) and (6) give the dependence of      
$\rho$ on $r$.  At  $r_{0}$, $\xi = 1$, and $\rho = \rho_{0}$
on both sides.  I.e.,  $\rho$ is continuous.  The value of $d\rho/dr$
is finite also but discontinuous at $r_{0}$, a fact already 
emphasized by Oliva[4].  So is  $d^{2}\rho/dr^{2}$.
 
\noindent {\bf 4.     Remarks}
 
The pseudopotential interaction [5] is , in q-number language:
\begin{equation}
V_{pp} = \frac{4\pi a \hbar^{2}}{m} \frac{1}{2} \int d {\vec{r}_{1}}
d {\vec{r}_{2}} \psi^{\dagger}({\vec{r}_{1}}) \psi^{\dagger}({\vec{r}_{2}}) 
\delta({\vec{r}_{1}} - {\vec{r}_{2}}) \frac{\partial}{\partial r_{12}} 
[r_{12} \psi(\vec{r}_{1})\psi(\vec{r}_{2})].
\end{equation}                                                              
                  
The commonly used delta-function interaction [5] is, also in 
q-number language,
\begin{equation}
V_{\delta} = \frac{4 \pi a \hbar^{2}}{m} \frac{1}{2} \int
d \vec{r} \psi^{\dagger} (\vec{r}) \psi^{\dagger} (\vec{r}) \psi(\vec{r})
\psi(\vec{r}).
\end{equation}
\noindent (a) We point out that there has been great confusion 
in the theoretical
literature of factors of two about $V_{pp}$ and $V_{\delta}$.  
Furthermore the
approximate one-particle c-number equation written down from (16) 
also has confusions of factors of two in the literature.
 
\noindent (b) Strictly speaking (16) does not make sense:  
The two-body interaction $A\delta(\vec{r}_{1} - \vec{r}_{2})$
 for $A < 0$ is not defined and for $A > 0$ is equivalent to zero.  
To prove
the first statement, we calculate the s-wave phase-shift for such a case.
The interaction is considered as the limit $R \rightarrow 0$
of a potential well of magnitude $U$ and radius $R$, keeping 
$UR^{3}$ = negative constant.  The wave function in the 
center-of-mass system at $R-$ is $R^{-1} \sin [{(-mU/\hbar^{2})}^{1/2}R]$.
The argument of sine approaches infinity as $R \rightarrow 0$.   To 
prove the second statement we do the same calculation and find all phase-
shifts to be zero.
 
\noindent (c)     Although (16) is strictly meaningless, 
the perturbation calculations
based on (15) and (16) in Refs.[5, 6], give the same result to 
the order $a^{1}$,
and is meaningful to that order.  The considerations of the present paper
are based on these results to order $a$  and are thus meaningful.
 
        How about higher orders?  This is a matter of 
some subtlety.  It was found
in Ref.[5] that to the second order in $a$, (16) gives infinity, 
confirming
its sickness, but that (15) gives a meaningful answer which 
is proportional to $a/L$  where $L$ is the size of the box.  To order    
$a^{3}$ (15) gives an energy containing a term equal to (constant) 
times $N^{3} a^{3}/L^{5}$ which approaches infinity 
as $L \rightarrow \infty$
while $\rho = N/L^{3}$ = constant.  This divergence was later 
removed [8] by a method of
summation over most divergent terms in a perturbation expansion:  A
summation starting with this divergent term leads to the convergent
expression $(4 \pi a N \rho) (128/15) {(a^{3}\rho/\pi)}^{1/2}$.
This result was
later confirmed by several authors who extended it to even higher 
orders.
 
        The method of Ref.[8] was further extended to 
cover finite temperatures
[9].  These and related developments had been summarized in a 
brief talk [10].
 
\noindent (d)     It is tempting to interpret (5) and (6) 
as indicating an additional
effective potential $V_{a} = kT 4 a \lambda^{2}\rho = 
8 \pi a \rho \hbar^{2}/m$,
and to interpret (14') as indicating an
additional potential of $V_{a} = 4 \pi a \rho \hbar^{2}/m$.
Such interpretations must be used with care.
In particular, it is not correct to assert that the condensate is in the
ground state of $V + V_{a}$.
 
        The work of CNY is supported in part by an NSF 
Grant PHY-9309888.  The work
of LHY is performed under the auspices of US DOE.

\newpage
\begin{center}
{\bf REFERENCES}
\end{center}
\begin{enumerate}
\item   M. H. Anderson, J. R. Ensher, M. R. Matthews, C. E. Wieman and E. A
.
Cornell, Science {\bf 269}, 198 (1995).
\item    C. C. Bradley, C. A. Sackett, J. J. Tollett and R. G. Hulet, Phys.
Rev. Lett. {\bf 75}, 1687 (1995).
\item    K. B. Davis, M. -O. Mewes, M. R. Andrews, N. J. van Druten, D. S.
Durfee, D. M. Kurn, and W. Ketterle, Phys. Rev. Lett. {\bf 75}, 
3969 (1995).
\item    There have been many theoretical papers on this subject.  See
references in Refs.[1-3] and V. V. Goldman, I. F. Silvera and A. J. Legget,
Phys. Rev. B{\bf 24}, 2870 (1981), J. Oliva, Phys. Rev. B{\bf 39}, 
4197 (1989) and others.
\item    K. Huang and C. N. Yang, Phys. Rev. {\bf 105}, 767 (1957).
\item    K. Huang, C. N. Yang and J. M. Luttinger, Phys. Rev. 
{\bf 105}, 776 (1957).
\item    R. P. Feynman, {\em Statistical Mechanics: a Set of 
Lectures} (Benjamin, New York, 1972).
\item   T. D. Lee, K. Huang and C. N. Yang, Phys. Rev. 
{\bf 106}, 1135 (1957).
\item   T. D. Lee and C. N. Yang, Phys. Rev. {\bf 112}, 1419 (1958).
\item   C. N. Yang, Physica {\bf 26}, S49 (1960).
\end{enumerate}
\vspace{0.25cm}
\begin{center}
{\bf FIGURE CAPTION} 
\end{center}
Fig. 1 \hspace{1cm} Example of  $\rho(r)$ as a function of  
$r$  for $a > 0$ according to Eq. (14').
The curve is for a harmonic trap.  Notice that the first and second
derivatives are both finite but discontinuous at $r_{0}$.
$\rho_{0} = \lambda^{-3} g_{3/2}(1)$ is the normal
part of $\rho$, and $\rho_{s}$ the condensate part.  
For smaller total number of particles $(=N)$, $r_{0}$ becomes smaller.  
It eventually shrinks to zero and the condensate disappears.
 
\end{document}